\documentstyle[preprint,prl,aps,epsf]{revtex}

\begin{document}
\preprint{LBNL-42170, August 12 1998} 

\title{Impact Parameter Dependence
of $J/\psi$ and Drell-Yan Production in Heavy Ion Collisions
at $\sqrt{s_{NN}} = 17.3$ GeV}

\author{V. Emel'yanov$^1$, A. Khodinov$^1$, S. R. Klein$^2$ and
R. Vogt$^{2,3}$}
 
\address{{
$^1$Moscow State Engineering Physics Institute (Technical
University), Kashirskoe ave. 31, Moscow, 115409, Russia\break 
$^2$Nuclear Science Division, Lawrence Berkeley National Laboratory, 
Berkeley, CA
94720, USA\break 
$^3$Physics Department, University of California, Davis, CA 95616, USA}\break}
 
\vskip .25 in
\maketitle
\begin{abstract}

In heavy-ion collisions,
$J/\psi$ and Drell-Yan production are expected to be
affected by nuclear modifications to the free nucleon structure
functions. If these modifications, known as shadowing, are proportional to the
local nuclear density, the per nucleon cross sections will depend on
centrality.  Differences in quark and gluon shadowing will lead to a new
source of impact parameter dependence of the $J/\psi$ to Drell-Yan production 
ratio.  We calculate this ratio in the CERN NA50 acceptance with several 
shadowing parameterizations to explore its centrality dependence.

\end{abstract}
\narrowtext

A significant `anomalous' suppression of $J/\psi$ production has been observed
in Pb+Pb collisions.  The ratio of
$J/\psi$ to Drell-Yan production is lower in central Pb+Pb
collisions than extrapolations from more peripheral collisions,
lighter ion interactions, and pA collisions suggest \cite{na50}.
Centrality is inferred from the transverse energy, $E_T$.

Almost all calculations of $J/\psi$ and Drell-Yan production in
nuclear collisions to date have been based on position independent
nucleon \cite{rvrev} or nuclear structure functions \cite{hammon}.
However, nuclear shadowing should depend on the parton's location
inside the nucleus.  If shadowing is due to gluon
recombination\cite{hot}, nuclear binding or rescaling \cite{close}, or
other local phenomena, it should be proportional to the local nuclear
density.  The only studies of spatial dependence have relied on
qualitative measurements of impact parameter, such as dark tracks in
emulsion\cite{E745}, to find evidence for such a spatial dependence.

This letter presents calculations of the impact parameter dependence
of nuclear shadowing on $J/\psi$ and Drell-Yan production in
heavy ion collisions.  This spatial dependence has an important effect
on the $E_T$ dependence of $J/\psi$ and Drell-Yan production.  Since
$J/\psi$ suppression is a predicted signature of quark-gluon plasma
formation \cite{MS}, interpretations \cite{rvrev} of the anomalous suppression
in the NA50 Pb+Pb data
\cite{na50} should include the spatial dependence
of the structure functions on the $J/\psi$ and Drell-Yan rates.  We focus
on the effects of shadowing and neglect $J/\psi$ absorption
mechanisms.

At leading order (LO), $J/\psi$ production is dominated by gluon
fusion while Drell-Yan production is due to $q\overline q$
annihilation.  Thus if the quark and gluon distributions are affected
differently by shadowing, and if shadowing has a spatial dependence, the
$J/\psi$ to Drell-Yan ratio can vary with impact parameter.
We present results in the acceptance of the CERN NA50 experiment \cite{na50}.

To be consistent with the NA50 analysis, we calculate $J/\psi$ and 
Drell-Yan production at leading order, LO. 
The LO cross
section for nuclei $A$ and $B$ colliding
at impact parameter $b$ with center of mass energy $\sqrt{s_{NN}}$
and producing 
a particle $V$ ($J/\psi$ or $\gamma^*$)
with mass $m$ at scale $Q$ is
\begin{eqnarray} \frac{d\sigma_{AB}^V}{dy dm^2
d^2b d^2r} = \sum_{i,j}\int \,dz \,dz'
F_i^A(x_1,Q^2,\vec{r},z) F_j^B(x_2,Q^2,\vec{b} - \vec{r},z')
\frac{d\widehat{\sigma}_{ij}^V
}{dy dm^2} \, \,  ,
\label{sigmajpsi}
\end{eqnarray}
where $\widehat{\sigma}_{ij}^V$ is the partonic $ij \rightarrow V$ 
cross section.
The nuclear parton densities, $F_i^A$, are
the product of momentum fraction, $x$, and $Q^2$ independent nuclear
densities, $\rho_A$; position and atomic mass, $A$, 
independent nucleon parton
densities, $f_i^N$; and a shadowing function, $S^i$:
\begin{eqnarray}
F_i^A(x,Q^2,\vec{r},z) = \rho_A(s) S^i(A,x,Q^2,\vec{r},z) 
f_i^N(x,Q^2) \, \, , 
\end{eqnarray}
where $s = \sqrt{r^2 + z^2}$.
In the absence of shadowing,
$S^i(A,x,Q^2,\vec{r},z)\equiv1$.  The nuclear density is given by a
Woods-Saxon distribution,
$\rho_A(s)= \rho_0 (1 + \omega (s/R_A)^2)/(1 + e^{(s-R_A)/d})$ 
where electron scattering data \cite{Vvv} are used to fix $R_A$, $d$, $\omega$
and $\rho_0$.

We use the color evaporation model of $J/\psi$ production \cite{HPC} so that
\begin{eqnarray}
\lefteqn{f_i^N(x_1,Q^2) f_j^N(x_2,Q^2) \frac{d\widehat{\sigma}_{ij}^{J/\psi}
}{dy dm^2} = K_{\rm th} \left\{
f_g^N(x_1,Q^2)f_g^N(x_2,Q^2)
\frac{\sigma_{gg}(m^2)}{m^2} \right.} \nonumber \\ & & \mbox{} \left.
+ \sum_{q=u,d,s} \big[f_q^N(x_1,Q^2) f_{\overline q}^N(x_2,Q^2) + f_{\overline 
q}^N(x_1,Q^2) f_q^N(x_2,Q^2)\big] 
\frac{\sigma_{q \overline q}(m^2)}{m^2}
\right\} \, \, .
\label{psi}
\end{eqnarray}
The LO partonic $c \overline c$ cross sections are defined in 
\cite{combridge} and $m^2= x_1x_2s_{NN}$. The 
fraction of $c \overline c$ pairs 
that become $J/\psi$'s is fixed at next-to-leading order (NLO) \cite{HPC}.
The ratio of the NLO to LO cross section is given by $K_{\rm th}$.  
We use the GRV LO \cite{GRV}
nucleon parton distributions with $m_c=1.3$ GeV and $Q=m_c$ as
well as the MRS A$^\prime$ densities \cite{mrsap} 
with $m_c=1.2$ GeV and $Q = 2m_c$ where $m_c$ and $Q$ are chosen to agree with 
data \cite{HPC}.    

The LO Drell-Yan cross section depends on isospin since
$\sigma_{pp}^{\rm DY} \neq \sigma_{pn}^{\rm DY} \neq \sigma_{np}^{\rm
DY} \neq \sigma_{nn}^{\rm DY}$
\begin{eqnarray}
\lefteqn{f_i^N(x_1,Q^2) f_j^N(x_2,Q^2) \frac{d\widehat{\sigma}_{ij}^{\rm DY}
}{dy dm^2} = K_{\rm exp} \frac{4 \pi \alpha^2}{9m^2s_{NN}}} \label{sigmady} \\ 
& & \mbox{} 
\times \sum_{q=u,d,s} e_q^2 \left[ \left\{{Z_A \over A} 
f_q^p(x_1,Q^2)+ {N_A\over A}
f_q^n(x_1,Q^2) \right\} \left\{ {Z_B \over B}
f_{\overline q}^p(x_2,Q^2) + {N_B\over B} 
f_{\overline q}^n(x_2,Q^2)\right\} + q 
\leftrightarrow \overline q \right] \nonumber \, \, ,
\end{eqnarray}
where $Z_A$ and $N_A$ are the number of protons and neutrons in the
nucleus.  We assume
that $f_u^p = f_d^n$, $f_d^p = f_u^n$ {\it etc.}  In Eqs.~(\ref{psi})
and (\ref{sigmady}), $x_{1,2} = Qe^{\pm y}/\sqrt{s_{NN}}$ and $Q = m$. The
factor $K_{\rm exp}$ accounts for the rate difference between the
calculations and the data.

Three parameterizations of shadowing, based on nuclear DIS 
\cite{Arn}, are used.  The first, $S_1(A,x)$,
treats quarks, antiquarks, and gluons
equally without $Q^2$ evolution \cite{EQC}. The other two evolve with $Q^2$ 
and conserve baryon number and total momentum.  The second, $S_2^i(A,x,Q^2)$,
modifies the valence quarks, 
sea quarks and gluons separately
for $2 < Q < 10$ GeV \cite{KJE}.  
The most recent, $S_3^i(A,x,Q^2)$, 
evolves each parton distribution 
separately for $Q \geq 1.5$ GeV \cite{KJEnew}.
The $S_3$ initial gluon distribution
shows important antishadowing in the region $0.1<x<0.3$ with sea quark 
shadowing in the same $x$ range.
In contrast, $S_2$ has less gluon antishadowing and essentially no sea quark 
effect.

Since we assume
that shadowing is proportional to
the local nuclear density, the spatial dependence is defined as
\begin{eqnarray}
S^i_{\rm WS} = 
S^i(A,x,Q^2,\vec{r},z) & = & 1 + N_{\rm WS}
[S^i(A,x,Q^2) - 1] \frac{\rho(s)}{\rho_0} \label{wsparam} \, \, ,
\end{eqnarray}
where $N_{\rm WS}=$ is chosen so that $(1/A)
\int d^3 s \rho(s) S^i_{\rm WS} = S^i$, similar to \cite{cast}. 
For lead, $N_{WS}=1.32$.
At large radii, $s \gg R_A$,
$S^i_{\rm WS} \rightarrow 1$ while at the nuclear center, 
the modifications
are larger than the average $S^i$.  
An alternative parameterization, $S_{\rm R}^i$,
proportional to the thickness of a spherical nucleus
\cite{us}, leads to a slightly larger modification in the nuclear core.

\def\rat{$\sigma^{J/\psi}/\sigma^{\rm DY}$}

We model our calculations to the NA50 experimental acceptance with
center of mass rapidity $0 <y_{\rm cm} < 1$ and decay angle in the
Collins-Soper frame $|\cos{\theta_{\rm CS}} | < 0.5$ \cite{na50}. The
Drell-Yan spectrum is measured for $m>4.2$ GeV and the factor $K_{\rm
exp}$ is extracted by comparing to Eq.~(\ref{sigmady}) calculated with
the GRV LO distributions \cite{GRV}. The ratio \rat\, is formed by
extrapolating the calculations to $2.9< m < 4.5$ GeV with the same
$K_{\rm exp}$ because $J/\psi$ and $\psi'$ decays dominate the region
$2.7 < m < 3.5$ GeV.  Table~\ref{psidytab} give the impact parameter
averaged cross sections per nucleon pair for the GRV LO and MRS
A$^\prime$ distributions, for $pp$, Pb+Pb and S+U interactions.  The
numbers show the effects of isospin and shadowing at $\sqrt{s_{NN}} = 17.3$
GeV.  With the MRS A$^\prime$ set, the isospin
correction is quite small in the extrapolated region
\cite{moriond}.  Because isospin is
unimportant in $J/\psi$ production, only the Pb+Pb cross section is
shown.  However, the $J/\psi$ data suggests that the $AB$
dependence might be stronger if shadowing could be removed from the
data.  Although the choice of parton densities influences the isospin
effect and $K_{\rm exp}$, the average Drell-Yan shadowing is changed
by less than 1\% while the difference in $J/\psi$ shadowing can be as
large as 5\%, largely because the gluon distribution is imprecisely
measured.  The table shows that the dependence on the nuclear species
is weak.  The change in the shadowing is also $\sim 1$\% when
$\sqrt{s_{NN}} = 19.4$ GeV.  Most important is the mass interval:
shadowing is 5\% stronger in the measured region than the extrapolated
region.

\begin{figure}[h]
\setlength{\epsfxsize=0.7\textwidth}
\setlength{\epsfysize=0.7\textheight}
\centerline{\epsffile{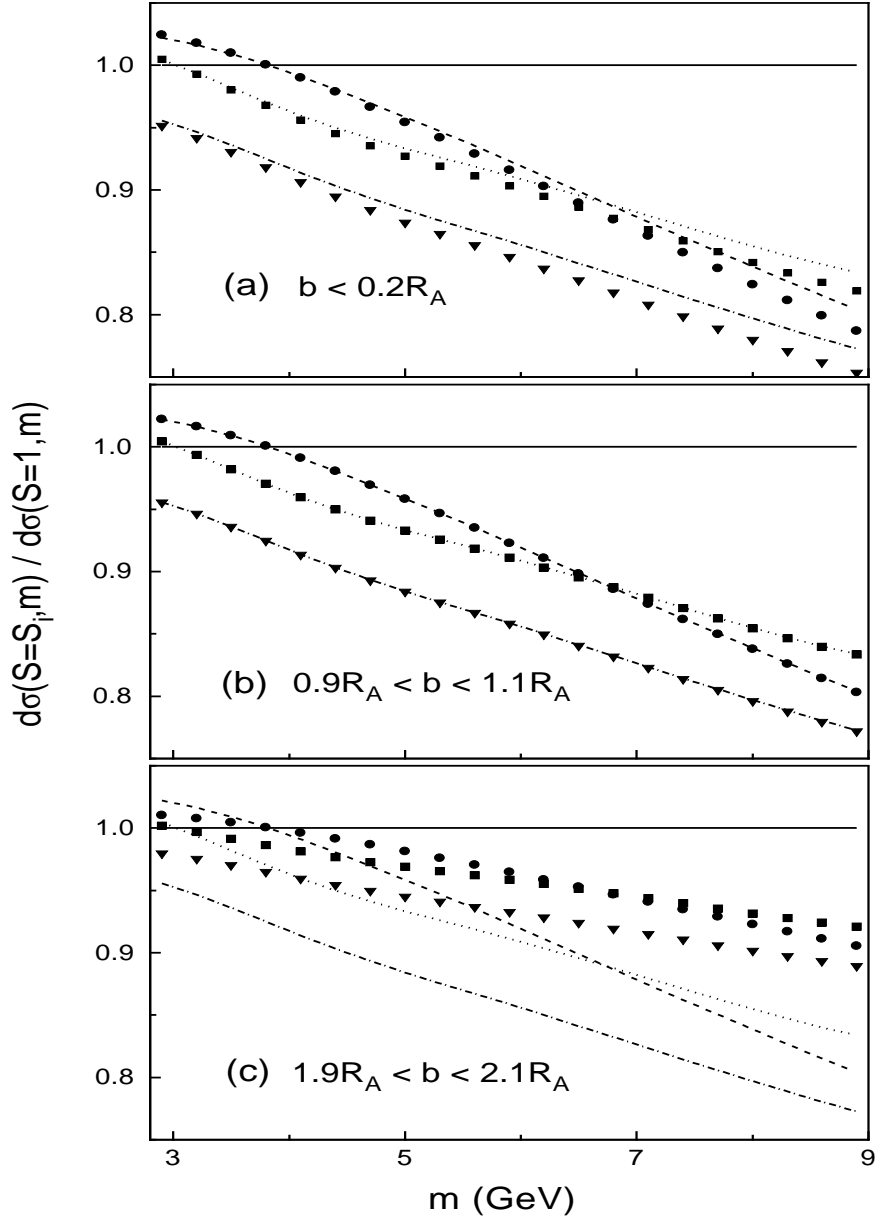}}
\caption[]{The Drell-Yan dilepton mass spectrum, $d\sigma^{\rm DY}/dM$, 
without and with spatial dependence relative to $S=1$.
The curves correspond to $b$-averaged results with
$S_1$ (dashed), $S_2$ (dotted) and $S_3$ (dot-dashed). The spatial dependence 
is illustrated for $S_{1, {\rm WS}}$ (circles)
$S_{2, {\rm WS}}$ (squares)
and $S_{3, {\rm WS}}$ (triangles). The impact parameter ranges are 
(a) $0<b<0.2R_A$ fm, (b) $0.9R_A<b<1.1R_A$ fm and (c) 
$1.9R_A<b<2.1R_A$ fm respectively.}
\label{fig1}
\end{figure}

To illustrate how shadowing could affect the interpretation of $J/\psi$
suppression, Fig.~\ref{fig1} shows the Drell-Yan mass distribution 
for $m>2.9$
GeV in three impact parameter bins relative to no shadowing, $S=1$.  
Shadowing changes the slope of the spectrum, producing a
$\sim 20$\% change in the predicted rate at $m \approx 9$ GeV.  
The absence of $Q^2$ evolution causes the $S_1$ results to decrease
faster with mass than $S_2$ or $S_3$.
Including only spatially-averaged shadowing increases $K_{\rm
exp}$ over that needed for $S=1$
in the measured region relative to the extrapolated region, 
shown in Table~\ref{psidytab}, as well as further increase the
discrepancy in central collisions while overestimating $K_{\rm exp}$
in peripheral collisions.  At small radii, Fig.~\ref{fig1}(a),
$d\sigma^{\rm DY}/dm$ drops more rapidly than the impact parameter
averaged spectra.  When $b \approx R_A$, Fig.~\ref{fig1}(b), the
averaged and spatial dependent spectra approximately coincide.  At
large radii, Fig.~\ref{fig1}(c), shadowing is reduced, approaching the
unshadowed spectra.  These results show that using a calculation to
extrapolate to an unmeasured region is problematic.

\begin{figure}[h]
\setlength{\epsfxsize=0.7\textwidth}
\setlength{\epsfysize=0.7\textheight}
\centerline{\epsffile{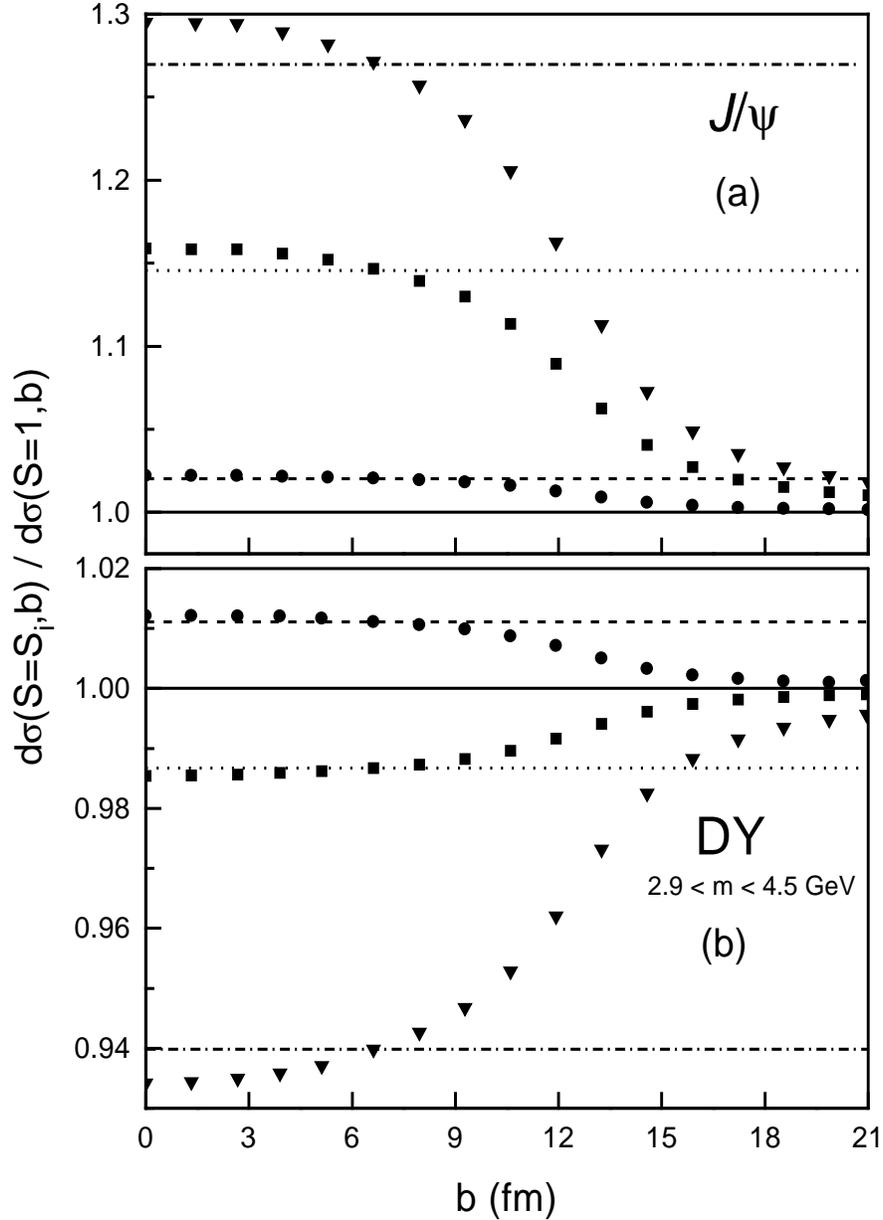}}
\caption[]{The (a) $J/\psi$ and (b) Drell-Yan rates in Pb+Pb collisions
at $\sqrt{s_{NN}} = 17.3$ GeV relative to
production without shadowing, $S=1$, as a function of impact parameter.
The curves and symbols are defined in Fig.~\ref{fig1}.}
\label{fig2}
\end{figure}

Figure~\ref{fig2} shows the impact parameter dependence of the
$J/\psi$ and Drell-Yan cross sections.  The $x_2$ ranges of the two
processes nearly coincide: the Drell-Yan region $2.9 <m< 4.5$ GeV
corresponds to $0.062<x_2<0.26$ while the $J/\psi$ cover
$0.066<x_2<0.18$.  Since all three parameterizations assume some gluon
antishadowing in this region, $\sigma^{J/\psi}$ is always enhanced in
Pb+Pb collisions.  Because $S_1$ is the same for quarks and gluons,
$J/\psi$ and Drell-Yan are equally affected by shadowing. On other
hand, $S_{2,3}^{\overline q} \leq 1$, reducing $\sigma^{\rm DY}$.

\begin{figure}[h]
\setlength{\epsfxsize=0.7\textwidth}
\setlength{\epsfysize=0.7\textheight}
\centerline{\epsffile{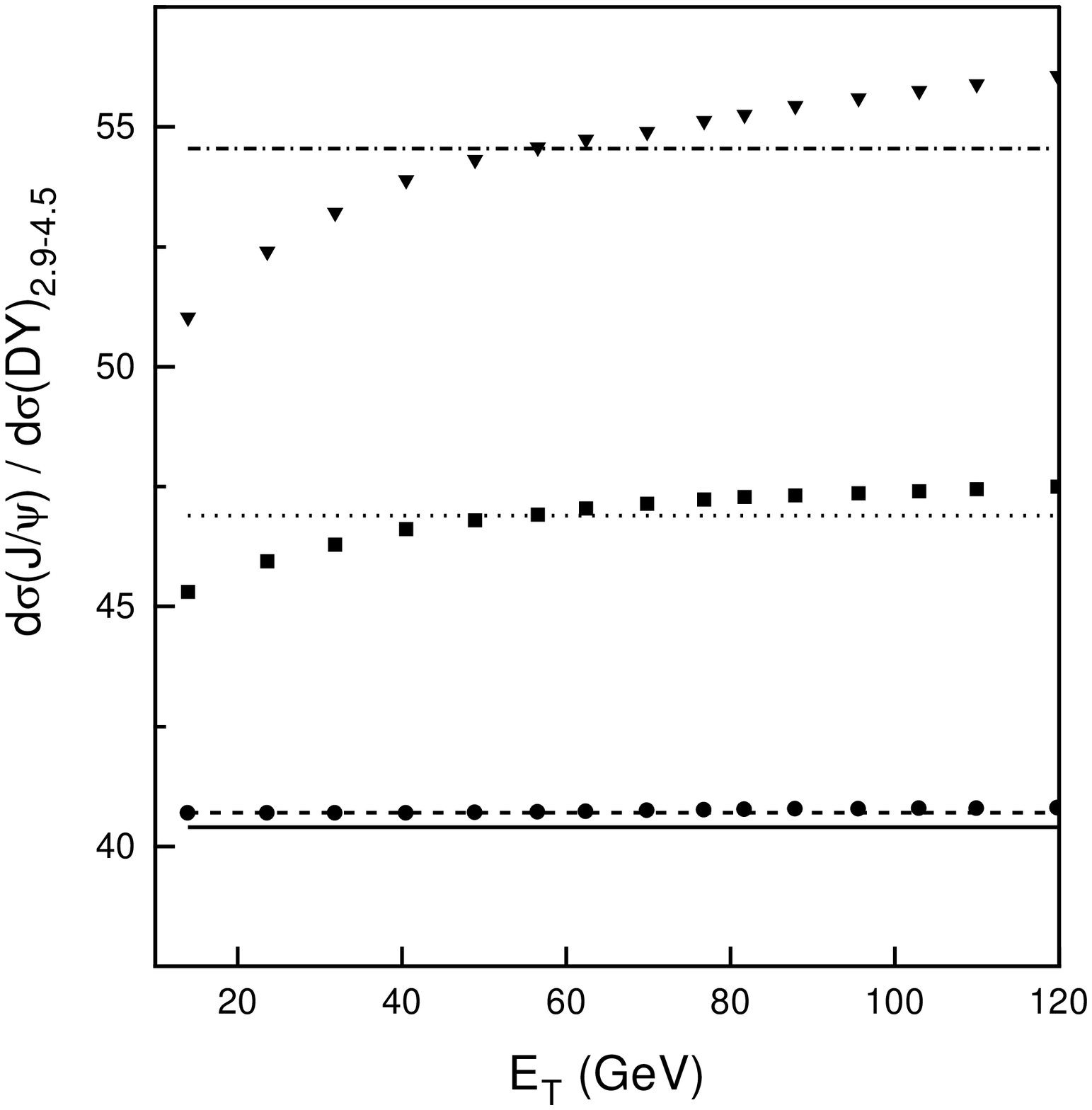}}
\caption[]{The ratio of $J/\psi$ to Drell-Yan production,
as a function of transverse energy, $E_T$.
The curves and symbols are defined in Fig.~\ref{fig1}.}
\label{fig3}
\end{figure}

In Fig.~\ref{fig3},
\rat, calculated in Eqs.~(\ref{sigmajpsi})-(\ref{sigmady}), is presented as a
function of $E_T$. The correlation between $E_T$ and $b$ 
is based on the number of nucleon participants \cite{rvrev},
in agreement with the most recent NA50 $E_T$ distributions
\cite{moriond}.  The Drell-Yan cross section is corrected for isospin
and \rat\, is calculated at $\sqrt{s_{NN}} = 19.4$ GeV.  After these
adjustments, with $S=1$, \rat $\sim 40.3$, in agreement with the
published NA50 data \cite{na50}.  The combined Drell-Yan shadowing
and $J/\psi$ antishadowing in Fig.~\ref{fig2} increases \rat to 40.6
with $S_1$, 44.5 for $S_2$ and 54.4 using $S_3$.  The $S_1$ ratio is
independent of $E_T$ because the $b$ dependence cancels.  However,
$S_2$ and $S_3$ vary with $E_T$.  The $S_2$ ratio rises about 7\%
while the $S_3$ ratio increases $\approx 11$\% as $\langle E_T\rangle
$ grows from 14 GeV to 120 GeV.  These enhancements are opposite to
the observed drop at large $E_T$ and small $b$ \cite{na50}, neglecting
shadowing.

Because of uncertainties in the gluon shadowing parameterization, 
it is difficult to draw detailed
conclusions.  However, $S_1$ should represent a 
lower limit and $S_3$ an upper limit.  A
stronger spatial dependence such as $S_{\rm R}^i$ would slightly increase the
effect with $E_T$ while parameterizations based
on {\it e.g.}\ nuclear binding \cite{close} might predict a
smaller effect. 

In conclusion, we have studied the impact parameter dependence of the ratio
\rat\, using a spatially dependent shadowing model.  We find
that the ratio increases at small $b$ (large $E_T$)
compared with more peripheral collisions.  The magnitude of the effect
depends on the chosen parameterization.  Neglecting shadowing could lead to an
increased $K_{\rm exp}$ at $\sqrt{s_{NN}} = 17.3$ GeV since the measured
cross section is more strongly 
affected by shadowing than the extrapolated cross section.
In addition, using an impact parameter averaged spectra in central
collisions would tend to underestimate the total number of Drell-Yan
pairs, increasing \rat\,.  
If this effect could be identified and corrected for
in the data, then \rat\, would rise $\sim 10$\% at low $E_T$ and drop
$\sim 4$\% at high $E_T$, enhancing the discrepancy between absorption models
and the data. At higher $\sqrt{s_{NN}}$, such as at future heavy-ion colliders, 
the shadowing effect will be
larger\cite{usinprep} 
since these colliders probe lower $x$ values. 

V.E. and A.K. would like to thank the LBNL RNC 
group for hospitality and M. Strikhanov and
V.V. Grushin for discussions and support.  R.V. would like to thank J. 
Schukraft for discussions.  We also thank K.J. Eskola
for providing the shadowing routines and for discussions.  This work
was supported in part by the Division of Nuclear Physics of the Office
of High Energy and Nuclear Physics of the U. S. Department of Energy
under Contract Number DE-AC03-76SF00098.

\begin{table}
\begin{tabular}{cccccccc}
  & \multicolumn{3}{c}{Drell-Yan ($2.9<m<4.5$ GeV)} & 
\multicolumn{3}{c}{Drell-Yan ($4.2<m<9$ GeV)}
& $J/\psi$ \\
  & $\sigma_{pp}$ (pb) & $\sigma_{\rm PbPb}$ (pb) &  $\sigma_{\rm SU}$ (pb)
  & $\sigma_{pp}$ (pb) & $\sigma_{\rm PbPb}$ (pb) &  $\sigma_{\rm SU}$ (pb)
  & $\sigma_{\rm PbPb}$ (nb) \\ \hline
\multicolumn{8}{c}{GRV LO} \\
$S=1$   & 16.6 & 12.7 & 13.3 & 2.31 & 1.66 & 1.76 & 1.61 \\
$S=S_1$ & $-$  & 12.8 & 13.5 & $-$  & 1.61 & 1.72 & 1.64 \\
$S=S_2$ & $-$  & 12.5 & 13.1 & $-$  & 1.56 & 1.67 & 1.84 \\
$S=S_3$ & $-$  & 11.9 & 12.6 & $-$  & 1.48 & 1.61 & 2.04 \\ \hline
\multicolumn{8}{c}{MRS A$^\prime$} \\
$S=1$   & 18.8 & 18.3 & 19.0 & 2.15 & 2.17 & 2.29 & 1.53 \\
$S=S_1$ & $-$  & 18.6 & 19.4 & $-$  & 2.10 & 2.26 & 1.58 \\
$S=S_2$ & $-$  & 18.1 & 18.9 & $-$  & 2.03 & 2.19 & 1.74 \\
$S=S_3$ & $-$  & 17.2 & 18.1 & $-$  & 1.93 & 2.10 & 1.85 \\ 
\end{tabular}
\caption[]{The Drell-Yan extrapolated and measured cross sections in
$pp$, Pb+Pb and S+U collisions and the $J/\psi$ cross section in Pb+Pb
collisions in the NA50 acceptance with the GRV LO and MRS A$^\prime$
parton densities at $\sqrt{s_{NN}} = 17.3$ GeV.  We have not included
$K_{\rm exp}=2.4$ for GRV LO and 1.7 for MRS A$^\prime$.}
\label{psidytab}
\end{table}

\begin{references}

\bibitem{na50} M.C. Abreu {\it et al.} (NA50 Collab.), Phys. Lett. {\bf B410},
337 (1997); Phys. Lett. {\bf B410}, 327 (1997).

\bibitem{rvrev} For a recent review, see R. Vogt, LBNL-41758 (1998),
to appear in Phys. Rep.

\bibitem{hammon} S. Gupta and H. Satz, Z. Phys. {\bf C55} 391, (1992); 
N. Hammon {\it et al.}, hep-ph/9807546.

\bibitem{hot} L.V. Gribov, E.M. Levin, and M.G. Ryskin, Phys. Rep. {\bf 100} 
1, (1983).

\bibitem{close} S. Kumano and F.E. Close, Phys. Rev. C {\bf 41}, 1855
(1990).

\bibitem{E745} T. Kitagaki {\it et al.}, Phys. Lett. {\bf B214}, 281
(1988).

\bibitem{MS} T. Matsui and H. Satz, Phys. Lett. {\bf B178} 416 (1986).

\bibitem{Vvv}
C.W. deJager, H. deVries, and C. deVries, Atomic Data and Nuclear Data 
Tables {\bf 14} 485, (1974).

\bibitem{HPC} R.V. Gavai {\it et al.}, Int. J. Mod. Phys. {\bf A10} 3043 
(1995); G.A. Schuler and R. Vogt,
Phys. Lett. {\bf B387} 181, (1996).

\bibitem{combridge}B.L. Combridge, Nucl. Phys. {\bf B151} 429, (1979).

\bibitem{hpdy}
S. Gavin {\it et al.}, Int. J. Mod. Phys. {\bf A10} 2961 (1995).

\bibitem{GRV} M. Gl\"{u}ck, E. Reya, and A. Vogt,
Z. Phys. {\bf C53} 127, (1992).

\bibitem{mrsap} A.D.~Martin, R.G.~Roberts and W.J. Stirling,
Phys. Lett. {\bf B354} 155, (1995).

\bibitem{Arn}  M. Arneodo, Phys. Rep. {\bf 240} 301
(1994); M.R. Adams {\it et al.}, Phys. Rev. Lett. {\bf 68} 
3266, (1992).

\bibitem{EQC}
K.J. Eskola, J. Qiu, and J. Czyzewski, private communication.

\bibitem{KJE}
K.J. Eskola, Nucl. Phys. {\bf B400} 240, (1993).

\bibitem{KJEnew}K.J. Eskola, V.J. Kolhinen and P.V. Ruuskanen,
hep-ph/9802350, Nucl. Phys. {\bf B} in press;
K.J. Eskola, V.J. Kolhinen and C.A. Salgado,
hep-ph/9807297.

\bibitem{cast} P. Castorina and A. Donnachie, Z. Phys. C {\bf 49},
481 (1991). 

\bibitem{us} V. Emel'yanov, A. Khodinov, S.R. Klein and R. Vogt,
Phys. Rev. Lett {\bf 81}, 1801 (1998); V. Emel'yanov,
A. Khodinov, S.R. Klein and R. Vogt, Phys. Rev. {\bf C56}, 2726
(1997); V. Emel'yanov, A. Khodinov and M. Strikhanov, Yad. Fiz. {\bf
60}, 539 (1997) [Phys. of Atomic Nuclei, {\bf 60} 465, (1997)];
 V. Emel'yanov and A. Khodinov, Yad. Fiz. {\bf
60}, 1489 (1997) [Phys. of Atomic Nuclei, {\bf 60} 1352, (1997)].

\bibitem{moriond}A. Romana {\it et al.} (NA50 Collab.), in Proceedings of
  the 33$^{\rm rd}$ Rencontres 
  de Moriond, {\em QCD and High Energy Hadronic Interactions}, Les
  Arcs, France, 1998.

\bibitem{usinprep}V. Emel'yanov, A. Khodinov, S.R. Klein and R. Vogt,
in preparation.
  
\end{references}
\end{document}